\documentclass[prd,amsmath,amssymb]{revtex4}

\usepackage{graphicx}
\usepackage{dcolumn}
\usepackage{bm}
\usepackage{cases}
\usepackage{mathtools}

\begin{document}

\title{Anomaly and Brownian fluid particle in Navier-Stokes turbulence\\ }


\author{Timo\hspace*{1mm}Aukusti\hspace*{1mm}Laine \vspace*{0.5cm}}
\email{timo.au.laine@gmail.com}



           
\begin{abstract}
We investigate the Navier-Stokes turbulence driven by a stochastic random Gaussian force. Using a field-theoretic approach, we uncover an anomaly that brings hidden structure to the theory. The anomaly is generated by a non-self-adjoint operator of the Jacobian and it follows the symmetries of the stochastic Navier-Stokes equation. We calculate the anomaly and demonstrate that by forcing the anomaly to vanish, the velocity field is constrained and a monopole-type object with a constant charge is formed. When the viscosity is zero, the anomaly can be interpreted as the Brownian damping coefficient of a random fluid particle. We provide the Brownian particle equation and its solution in the presence of a pump and viscosity. Our results suggest that the anomaly is an inherent feature of stochastic turbulence and must be taken into account in all stochastic turbulence calculations. This constitutes an additional law for the original set of stochastic Navier-Stokes equations.
\end{abstract}

\maketitle

\vspace*{0.5cm}

\section{\label{sec0}Introduction}

Navier-Stokes turbulence is a fascinating and complex phenomenon that represents one of the most challenging and elusive problems in classical physics. It is a branch of fluid dynamics that seeks to clarify the chaotic and turbulent motion of fluids, which is characterized by rapid and irregular swirling of fluid particles. These phenomena are typically described by the Navier-Stokes equations, partial differential equations that govern the conservation of momentum and mass in fluid flow. Although these equations have helped to understand turbulence, due to their nonlinearity, they are still an active area of research in flow dynamics.

Navier-Stokes turbulence within field theory represents an intersection of flow dynamics and theoretical physics. This approach extends the traditional understanding of the Navier-Stokes equations to field theory and brings a new perspective to the behavior of turbulent flows. It allows us to dive into the complex interaction of various physical quantities in a turbulent flow, from velocity and pressure fields to vortices and energy distribution. This unique perspective is invaluable in understanding the statistical properties of turbulence that often elude traditional approaches. The application of field theory to turbulence brings a new perspective, where fluid properties are treated as fields and an attempt is made to understand their behavior through quantum field theory, statistical mechanics or other field theoretical approaches, Refs.~\onlinecite{frisch}-\onlinecite{ramond}.

In this article we will explore the stochastic Navier-Stokes turbulence and extend the results of Burgers turbulence to cover the Navier-Stokes equation, Ref.~\onlinecite{laine}.

\section{\label{sec1}Navier-Stokes turbulence}

In this section, we derive the generating functional of correlation functions based on the stochastic Navier-Stokes equation. This is a basis function that can be used to calculate various types of correlations in turbulent motion, Ref.~\onlinecite{zinn}.\\

\noindent
{\bf Notations}\\

We explore the following stochastic Navier-Stokes equations 

\begin{eqnarray}
  {\rm A)} && \frac{\partial {\vec U}}{\partial t}+ ({\vec U} \cdot \nabla) {\vec U}  = 
   \frac{1}{\rho}\nabla p
  +\nu \nabla^2 {\vec U} +{\vec f}, \label{eq1}\\ [1ex]
  {\rm B)} && \nabla\cdot \vec{U} = 0, \label{eq1b}
\end{eqnarray}

\noindent
where  

\begin{equation}
  \vec{U} = {\vec U}(t,\vec{x}) = u_1 \hat{i} + u_2\hat{j}+u_3 \hat{k}
\end{equation}

\noindent
is the velocity field, $\rho$ is the mass density, $p$ is the pressure, $\nu$ is the viscosity, and $\vec{f}=\vec{f}(t,\vec{x})$  is the Gaussian random force satisfying the condition

\begin{equation}
  \langle f_i(t,{\vec x})f_j(t',{\vec y}) \rangle = \kappa({\vec x}-{\vec y})\delta_{ij}\delta(t-t'). \label{eq2}
\end{equation}

\noindent
Function $\kappa(\vec{x}-\vec{y})$ defines the spatial correlation of the random forces. While the density is constant, the fluid is assumed incompressible (\ref{eq1b}).
We introduce the notation

\begin{equation}
  x_{\mu}\hspace*{0.5cm}\mu = 0,1,2,3 \hspace*{0.2cm}{\rm with}
  \hspace*{0.2cm}x_0=t,\hspace*{0.3cm}(x_1,x_2,x_3)=(x_i) = \vec{x},
\end{equation}

\noindent
i.e.

\begin{equation}
  x_{\mu} = (x_0,x_i) = (t,\vec{x}), \hspace*{0.3cm}i=1,2,3.
\end{equation}

\noindent
We write equation (\ref{eq1})  as

\begin{equation}
  \partial_t u_i+ (u_j \partial_j) u_i  =
  \partial_i \tilde{p}+
   \nu \partial_j \partial_j  u_i + f_i, \label{eq01}
\end{equation}

\noindent
where $\tilde{p} = p/\rho$, $\partial_j = \partial/\partial x_j$ and $i = 1,2,3$. 
We have also used a shortened notation

\begin{equation}
  {\vec U} \cdot \nabla = \sum_{j=1}^{3} u_j\partial_j = u_j\partial_j ,
\end{equation}

\noindent
where repeated indices are summed expect when otherwise indicated.\\

\noindent
{\bf Generating functional}\\

The generating functional of correlation functions is defines as, Ref.~\onlinecite{zinn},

\begin{equation}
  \langle F[\lambda]\rangle = \int Df Du F[\lambda] \exp(-S[f,u]).  \label{eq:S00}
\end{equation}

\noindent
The $ \langle F[\lambda]\rangle$ can be written with the help of $\delta$-functions (when $i=1$),

\begin{eqnarray}
   &&\int D f_1Du_1 
   \delta( \partial_t u_1+ (u_j \partial_j) u_1 -
  \partial_1 \tilde{p}-
   \nu \partial_j \partial_j  u_1- f_1 )J[u]
   \exp \Bigl (-\frac{1}{2}\int dtd^3xd^3y f_1(t,\vec{x}) \kappa^{-1}(\vec{x}-\vec{y})f_1(t,\vec{y})\Bigr )\\
   &&
   = \int Du_1 D\mu_1
   J[u]
   \exp \Bigl (-\frac{1}{2}\int dtd^3xd^3y \mu_1(t,\vec{x}) \kappa(\vec{x}-\vec{y})\mu_1(t,\vec{y})
   +i\int d^4x \mu_1( \partial_t u_1+ (u_j \partial_j) u_1 -
  \partial_1 \tilde{p}-
   \nu \partial_j \partial_j  u_1 )
   \Bigr ) \nonumber
   .  \label{eq:S01}
\end{eqnarray}

\noindent
Here, $J[u]$ is the Jacobian or determinant of the function inside the delta function. Since $\vec{U}$ is a vector, we could write 3 similar equations for each axis $1,2,3$. Alternatively, we introduce a matrix or metric $h^{ij}$ which is zero except for $h^{11} = \hat{i}$ , $h^{22} = \hat{j}$ , and $ h ^ { 33} = \hat{k}$ and has the property

\begin{equation}
\mu^i = h^{ij}\mu_j.
\end{equation}

\noindent
Also, for example, 

\begin{equation}
   \mu^i f_i = \mu_1 f_1 \hat{i}+\mu_2 f_2 \hat{j}+\mu_3 f_3 \hat{k}.
\end{equation}

\noindent
Then we may write the generating functional as

\begin{equation}
  \langle F[\lambda]\rangle = \int Du D\mu F[\lambda] J[u]\exp(-S[u,\mu]),  \label{eq:S0}
\end{equation}

\noindent
where the action $S[u(t,\vec{x}),\mu(t,\vec{x})]$ is

\begin{equation}
  S[u,\mu] = \frac{1}{2} \int dtd^3 xd^3y \mu^i(t,{\vec x})\kappa({\vec x}-{\vec y})\mu_i(t,{\vec y}) -i\int
d^4x \mu^i(\partial_t u_i+u_j\partial_j u_i
-\partial_i \tilde{p}-\nu\partial_j\partial_j u_i
).  \label{eq:S}
\end{equation}

\noindent
$S$ is a kind of vector, but it can be understood as consisting of 3 different equations, all of which are described by one equation. \\

\noindent
{\bf Jacobian}\\

We determine the Jacobian $J[u]$. It is not necessarily obvious what the functional derivative of a vector is. Usually a functional derivative is applied to the scalar field. That's why we use the definition

\begin{equation}
 \vec{U} = u\vec{e},
\end{equation}

\noindent
where $\vec{e} = e_1\hat{i}+e_2\hat{j}+e_3\hat{k}$ and write the Jacobian of the stochastic Navier-Stokes equation as

\begin{equation}
  J[u] = \det \biggl | \frac{\delta \vec{f}}{\delta u}\biggr | = \frac{\delta}{\delta u}
  \Bigl ( \frac{\partial u}{\partial t}+ u(e_j\partial_j) u  
  -\nu \partial_j\partial_j u \Bigr )\vec{e}.
\end{equation}

\noindent
We have dropped the pressure term because it does not depend on $u$. The Jacobian is also a vector function consisting of a determinant and a unit vector $\vec{e}$.
The functional derivative can now be calculated and the determinant becomes

\begin{equation}
  J[u]  =
  \det | \partial_t+(e_j\partial_j)u+u(e_j\partial_j)-\nu \partial_j\partial_j | \vec{e}.
\label{eq:jac}
\end{equation}

\noindent
 The value of the determinant is the same for all three directions. By using anticommuting functions, $\Psi = \Psi(t,\vec{x})$ and $\tilde{\Psi} = \tilde{\Psi}(t,\vec{x})$, Refs.~\onlinecite{zinn}-\onlinecite{ramond},  the determinant can be written as

\begin{equation}
  J[u] = \int D\bar{\Psi}D\Psi \exp(-S_A),
\end{equation}

\noindent
where the action of the determinant is

\begin{equation}
  S_A = -\vec{e}\int d^4x \bar{\Psi}(\partial_t+(e_j\partial_j)u+u(e_j\partial_j)-\nu \partial_j\partial_j )\Psi. \label{det1}
\end{equation}

\noindent
The action is also a vector with 3 components.\\

\noindent
{\bf Full action}\\

\noindent
We collect the results. Action (\ref{eq:S}) together with the Jacobian (\ref{det1}) form the complete action,

\begin{eqnarray}
  S_{full}[u,\mu,\Psi,\tilde{\Psi}] =&& \frac{1}{2} \int dtd^3 xd^3y \mu^i(t,{\vec x})\kappa({\vec x}-{\vec y})\mu_i(t,{\vec y}) -i\int
d^4x \mu^i(\partial_t u_i+u_j\partial_j u_i
-\partial_i \tilde{p}-\nu\partial_j\partial_j u_i
) \nonumber \\
&& - \vec{e}
\int d^4x \bar{\Psi}(\partial_t+(e_j\partial_j)u+u(e_j\partial_j)-\nu \partial_j\partial_j )\Psi. \label{eq:S2}
\end{eqnarray}

\noindent
 The action (\ref{eq:S2}) has all the information of the stochastic Navier-Stokes equation in one equation. \\

\section{\label{sec2}Effective action when viscosity is zero}

In this section we calculate the determinant or Jacobian (\ref{eq:jac}). First, it should be noted that the determinant depends on the velocity field $u$

\begin{equation}
  \frac{\delta}{\delta u} S_A = \vec{e}e_j\partial_j \bar{\Psi}\Psi \neq  0,
\end{equation}

\noindent
so it cannot be ignored.
The problem is that the operator in the determinant is non-self-adjoint, and if we want to express the action without the Jacobi fields, i.e. using only $u$, we need to calculate the square of the determinant. This is analogous to chiral field theories. Therefore we first need to define the "right-handed" and "left-handed" actions or determinants after which the square can be calculated.
\\

\noindent
{\bf Chiral determinants}\\

We observe that

\begin{equation}
  \int d^4x \bar{\Psi}(\partial_t+(e_j\partial_j)u+u(e_j\partial_j)-\nu \partial_j\partial_j )\Psi
  = \int d^4x (-\partial_t-u(e_j\partial_j)-\nu \partial_j\partial_j )\bar{\Psi}\Psi. \label{det123}
\end{equation}

\noindent
We get the right-handed determinant, which is

\begin{equation}
  S_R = - \vec{e}\int d^4x \bar{\Psi}(\partial_t+(e_j\partial_j)u+u(e_j\partial_j)-\nu \partial_j\partial_j )\Psi, \label{det1a}
\end{equation}

\noindent
and the left-handed determinant is

\begin{equation}
  S_L = \vec{e}\int d^4x (-\partial_t-u(e_j\partial_j)-\nu \partial_j\partial_j )\bar{\Psi}\Psi
  = -\vec{e}\int d^4x \Psi(-\partial_t-u(e_j\partial_j)-\nu \partial_j\partial_j )\bar{\Psi}. \label{det1b}
\end{equation}

\noindent
An additional "$-$" sign comes from the integral,

\begin{equation}
 \int D\bar{\Psi}D\Psi \exp^{S[\bar{\Psi},\Psi]} = -\int D\Psi D\bar{\Psi}\exp^{-S[\Psi,\bar{\Psi}]} 
 = \int D\Psi D\bar{\Psi}\exp^{S[\Psi,\bar{\Psi}]} .
\end{equation}

\noindent
The square of the chiral determinant is

\begin{equation}
 \det | \partial_t+(e_j\partial_j)u+u(e_j\partial_j)-\nu \partial_j\partial_j | = \Bigl [
  \det | \partial_t+(e_j\partial_j)u+u(e_j\partial_j)-\nu \partial_j\partial_j | 
   \det | -\partial_t-u(e_j\partial_j)-\nu \partial_j\partial_j | \Bigr ]^{1/2}. \label{sq22}
\end{equation}

\noindent
The causality needs to be preserved when calculating the determinant.
\\

\noindent
{\bf Square of the determinant} \\

First we set the viscosity to zero. The challenge of calculating the square of the determinant (\ref{sq22}) is that the usual methods of two-dimensional chiral methods are not applicable. They use complex values and vectors, but now all is real. Alternatively, we require that $S_R = S_L$ and use the definition of the determinant, which is
the determinant is the product of its eigenvalues. Therefore, we consider the following eigenvalue equations

\begin{equation}
  \begin{cases}
(\partial_t+(e_j\partial_j)u+u(e_j\partial_j))A=\lambda A, \\ 
(-\partial_t-u(e_j\partial_j))A=\lambda A,
  \end{cases} \label{eigeneq}
\end{equation}

\noindent
where $\lambda = \lambda(t,\vec{x},u)$ is an eigenvalue and $A = A(t,\vec{x},u)$ is the corresponding eigenfunction. For consistency, both determinants must produce the same eigenvalue functions. Putting the equations (\ref{eigeneq}) together gives

\begin{equation}
  \lambda = \frac{1}{2}(e_j\partial_j)u(t,\vec{x}). \label{eigenval}
\end{equation} 

\noindent
By analogy with chiral theories, the result is already the "square root". The determinant with the Jacobi fields is then

\begin{equation}
   S_{D}(\nu = 0) =   - \vec{e} \int d^4x \bar{\Psi}\frac{1}{2}(e_j\partial_j )u(t,\vec{x})\Psi. \label{det20100023}
\end{equation}

\noindent
One can use symmetries to verify that the result is correct, Ref.~\onlinecite{laine}. An alternative way to derive the anomaly is to note that

\begin{equation}
  \frac{\delta}{\delta u} S_D =  \frac{\delta}{\delta u} S_R +  \frac{\delta}{\delta u} S_L.
\end{equation}

\noindent
As the definition of the determinant states, the value of the determinant is then 

\begin{equation}
  \det | \partial_t+(e_j\partial_j)u+u(e_j\partial_j) | = \frac{1}{2}(e_j\partial_j)u(t,\vec{x}).
\label{eq:jac3}
\end{equation}

\noindent
The determinant cannot be zero, because then the generating functional (\ref{eq:S0}) is zero and we have no physics (or some type of regularization is needed to treat the divergence).
 We write the determinant as a path-integral. By using the identity 

\begin{equation}
 {\rm det M} = \exp {{\rm Tr} [ {\rm ln M} } ],
\end{equation}

\noindent
where M is a matrix, we may write

\begin{equation}
    {\rm det M} = \exp(-S_D),
\end{equation}

\noindent
and

\begin{equation}
   S_{D}(\nu = 0) =    -\vec{e} \int d^4x \ln (\frac{1}{2}e_j\partial_j u(t,\vec{x})). \label{det201000}
\end{equation}

\noindent
This is a path-integral representation of the square root of the determinant. 
The anomaly is due to the second term of the stochastic Navier-Stokes equation (\ref{eq1}). This is consistent with the observation that

\begin{equation}
  \vec{\mu}\cdot(\vec{U}\cdot\nabla)\vec{U}
\end{equation}

\noindent
can also be understood as the interaction Lagrangian of a charged particle with a Dirac monopole or a magnetic point vortex, Ref.~\onlinecite{jackiw}.
\\

\noindent
{\bf Constraint} \\

In path-integrals, local symmetry transformations are particularly interesting because they describe physical phenomena typically observed in nature. We have shown that in general the determinant of the stochastic Navier-Stokes equation depends on the velocity field (\ref{eq:jac3}) and it cannot be neglected. However, one can consider situations where the anomaly is forced to vanish, i.e. the determinant is required to be a number which cause constraints on the field $u$. 
The requirement of the determinant to be constant gives the condition

\begin{equation}
  \det | \partial_t+(e_j\partial_j)u+u(e_j\partial_j) | = \frac{1}{2}e_j\partial_j u(t,\vec{x})
  = {\rm constant}.
\label{eq:jac3212}
\end{equation}

\noindent
We call this "fixing the gauge", i.e. we have chosen the determinant to be a constant number. We could select some other gauge as well. Later we will show that if we want to examine the symmetries of the stochastic Navier-Stokes turbulence equation, this gauge is the only correct choice.

From (\ref{eq:jac3212}) we derive the constraint condition

\begin{equation}
 \frac{\delta }{\delta u} {\rm det } = 0,
\end{equation}

\noindent
or

\begin{equation}
(e_j\partial_j)\delta u(t,\vec{x}) =0. \label{condiss}
\end{equation}

\noindent
These are the conditions we expect from the anomaly. 
The variation of the function (\ref{det201000}) is

\begin{equation}
   \delta S_D =   -2\vec{e}\int d^4x \frac{1}{e_i\partial_i u(t,\vec{x})}
   e_j\partial_j \delta u(t,\vec{x}). \label{consti8}
\end{equation}

\noindent
This integral is zero if

\begin{eqnarray}
&& \delta u = {\rm constant}, \hspace*{0.5cm}{\rm or}\\
&&e_j\partial_j \delta u=0,\hspace*{0.5cm}{\rm or} \label{constraint2} \\
&& e_j\partial_je_i\partial_iu=0.
\end{eqnarray}

\noindent
The equations show the constraint correctly.
\\

\noindent
{\bf Zero eigenvalue} \\

Zero eigenvalues are especially interesting because they typically enable some additional physical property or observable. In the stochastic Navier-Stokes turbulence there is a zero eigenvalue in the determinant.
We have the eigenvalue equation (\ref{eigeneq})

\begin{equation}
  (\partial_t+(e_j\partial_j)u+u(e_j\partial_j))A(t,\vec{x},u)=\lambda A(t,\vec{x},u)
  = \frac{1}{2}(e_j\partial_j)u(t,\vec{x})A(t,\vec{x},u). \label{eigeq0}
\end{equation}

\noindent
From (\ref{eigeq0}) we get the zero eigenvalue equation  

\begin{equation}
   (\partial_t+\frac{1}{2}e_j\partial_j u+u e_j\partial_j)A(t,\vec{x},u)=0\cdot A(t,\vec{x},u)=0
. \label{eigeq1}
\end{equation}

\noindent
We note that the constraint imposes no restrictions on $t$. Therefore we have

\begin{eqnarray}
  &&\partial_t A(t,\vec{x},u) =  0, \hspace*{0.5cm}{\rm and}\label{condit1}\\
  &&u e_j\partial_jA(t,\vec{x},u) =  -\frac{1}{2}e_j \partial_juA(t,\vec{x},u).  \label{condit2} 
\end{eqnarray}

\noindent
 We will use these values later when calculating the viscosity of the anomaly. \\

\noindent
{\bf Effective action} \\

We collect the results when the viscosity is zero. The effective action is

 \begin{equation}
  S_{eff}[\nu=0] = \frac{1}{2} \int dtd^3 xd^3y \mu^i(t,{\vec x})\kappa({\vec x}-{\vec y})\mu_i(t,{\vec y}) -i\int
d^4x \mu^i(\partial_t u_i+u_j\partial_j u_i
-\partial_i \tilde{p}
) 
 - \vec{e}
\int d^4x \bar{\Psi}\frac{1}{2} e_j\partial_j u\Psi, \label{acteff11000}
\end{equation}
 
\noindent
or

\begin{equation}
 S_{eff}[\nu=0] = \frac{1}{2} \int dtd^3 xd^3y \mu^i(t,{\vec x})\kappa({\vec x}-{\vec y})\mu_i(t,{\vec y}) -i\int
d^4x \mu^i(\partial_t u_i+u_j\partial_j u_i
-\partial_i \tilde{p}
)  -  \vec{e}\int d^4x \ln (   \frac{1}{2}e_j\partial_j u(t,x)). \label{acteff110}
\end{equation}

\noindent
The constraint

\begin{equation}
 e_j\partial_j \delta u(t,\vec{x}) = 0   \label{constr2}
\end{equation} 

\noindent
 is a condition for which the anomaly vanishes. Equation (\ref{acteff11000}) is the same as the equation (\ref{eq:S2}), but in (\ref{acteff11000}) the determinant is a square root, which can then be expressed in terms of the field $u$, and this is the equation (\ref{acteff110}).\\

\section{\label{sec3}Effective action when viscosity is non-zero }

Next, we consider the situation where the viscosity is non-zero. Again, we need to calculate the square root of the determinant. The action with the Jacobi fields is 

 \begin{eqnarray}
  S_{eff} &=& \frac{1}{2} \int dtd^3 xd^3y \mu^i(t,{\vec x})\kappa({\vec x}-{\vec y})\mu_i(t,{\vec y}) -i\int
d^4x \mu^i(\partial_t u_i+u_j\partial_j u_i
-\partial_i \tilde{p} - \nu\partial_j\partial_j u_i
)  \nonumber\\ &&
 - \vec{e}
\int d^4x \bar{\Psi}(\frac{1}{2} e_j\partial_j u-\nu\partial_j\partial_j)\Psi. \label{acteff110000}
\end{eqnarray}

\noindent
We calculate the determinant. 
We use the following set of eigenvalue equations

\begin{equation}
  \begin{cases}
(\partial_t+(e_j\partial_j)u+u(e_j\partial_j)-\nu\partial_j\partial_j)A'=\lambda' A', \\ 
(-\partial_t-u(e_j\partial_j)-\nu\partial_j\partial_j)A'=\lambda' A',
  \end{cases} \label{cases1}
\end{equation}

\noindent
where $\lambda ' = \lambda '(t,\vec{x},u)$ is the eigenvalue and $A '= A'(t,\vec{x},u)$ is the corresponding eigenfunction. This gives the result 

\begin{equation}
  \lambda ' = \frac{1}{2} e_j\partial_j u(t,\vec{x}) - \nu \frac{\partial_j\partial_jA'}{A'}
  = \lambda - \nu \frac{\partial_j\partial_jA'}{A'},
  \label{visc1}
\end{equation} 

\noindent
where $\lambda$ is the eigenvalue of the non-viscosity equations. We add (\ref{visc1}) back into the equation (\ref{cases1}). This gives

\begin{equation}
  \begin{cases}
(\partial_t+(e_j\partial_j)u+u(e_j\partial_j))A'=\lambda A', \\ 
(-\partial_t-u(e_j\partial_j))A'=\lambda A',
  \end{cases}  \label{eigeneq11}
\end{equation}

\noindent
 from which it follows that $A'=A$, and $A$ is the eigenvector of the non-viscosity equations. This means that even in the presence of viscosity, the viscosity does not couple to or change the eigenvector. We have then

 \begin{equation}
  \lambda '
  = \frac{1}{2}e_j\partial_j u - \nu \frac{\partial_j\partial_jA}{A} .
\label{eq:jac5}
\end{equation}

\noindent
The condition (\ref{condit2}) can be solved with respect to $A$

\begin{equation}
  A(t,\vec{x},u) = Au(t,\vec{x})^{-1/2}.
\end{equation}

\noindent
The eigenvalue $\lambda '$ can be calculated

\begin{equation}
  \lambda '= \frac{1}{2}e_j\partial_j u -\nu F[u] =
  \frac{1}{2}e_j\partial_j u -\nu \Bigl (\frac{3}{4}\frac{(\partial_ju)^2}{u^2} -\frac{\partial_j\partial_j u}{2u} \Bigr ).
\label{eq:jac512}
\end{equation}
 
 \noindent
 The Jacobian is

\begin{equation}
    J[u] = \exp \Bigl (  \vec{e}\int d^4x \ln ( \frac{1}{2}e_j\partial_j u -\nu \Bigl (\frac{3}{4}\frac{(\partial_ju)^2}{u^2} -\frac{\partial_j\partial_j u}{2u} \Bigr ) )\Bigr ), \label{det20}
\end{equation}

\noindent
and the effective action then becomes 

\begin{eqnarray}
 S_{eff} 
&=& \frac{1}{2} \int dtd^4 xd^3y \mu^i(t,{\vec x})\kappa({\vec x}-{\vec y})\mu_i(t,{\vec y}) -i\int
d^4x \mu^i(\partial_t u_i+u_j\partial_j u_i
-\partial_i \tilde{p}-\nu\partial_j\partial_j u_i
) \nonumber \\
&&- \vec{e}\int d^4x \ln ( \frac{1}{2}e_j\partial_j u -\nu \Bigl (\frac{3}{4}\frac{(\partial_ju)^2}{u^2} -\frac{\partial_j\partial_j u}{2u} \Bigr ) )\ .
\label{acteff1}
\end{eqnarray}

\noindent
When the velocity field is complex differentiable in an open set, it can be shown by the power series that

\begin{equation}
  F[u] = \frac{3}{4}\frac{(\partial_ju)^2}{u^2} -\frac{\partial_j\partial_j u}{2u} > 0.
\end{equation}

\noindent
The viscosity becomes part of the constraint.\\

\section{\label{sec31}Brownian particle }

When the viscosity is zero, the action integral can be written without the anomaly term, i.e. by forcing the anomaly to vanish.
We reformulate the action of the stochastic Navier-Stokes equation so that it does not contain the anomaly.  \\

\noindent
{\bf Notations} \\

We define an object $L$,

\begin{equation}
  L = L[u] = e_j \partial_j u.
\end{equation}

\noindent
$L$ has the property

\begin{equation}
  L = {\rm constant} \neq 0, \label{L1}
\end{equation}

\noindent
and it does not change as $u$ varies

\begin{equation}
  \frac{\delta}{\delta u} L =0,\hspace*{0.5cm}{\rm or}\hspace*{0.5cm} \delta L = 0.\label{L2}
\end{equation}

\noindent
Equations (\ref{L1}) and (\ref{L2}) are the anomaly constraint conditions. Additionally, $L$ has symmetry properties

\begin{equation}
  \tilde{L}[\tilde{t},\tilde{\vec{x}},\tilde{u}] = f_1(t,\vec{x},u)L[f_2(t,\vec{x},u)],
\end{equation}

\noindent
where $f_1$ and $f_2$ are some functions.
These are shown in the next section.\\

\noindent
{\bf Fluid particle} \\

We may now write

\begin{equation}
  (\vec{U}\cdot\nabla)\vec{U} = ue_j\partial_ju\vec{e} = uL\vec{e} = L\vec{U}. \label{eqdamp}
\end{equation}

\noindent
Note that the equation (\ref{eqdamp}) can be interpreted as an eigenvalue equation $\vec{U}\cdot\nabla$ with eigenvalue $L$ and eigenvector $\vec{U}$.
The stochastic Navier-Stokes equation (\ref{eq1}) becomes

\begin{equation}
  \frac{\partial {\vec U}}{\partial t}+ L\vec{U} = 
   \frac{1}{\rho}\nabla p
  +\nu \nabla^2 {\vec U} +{\vec f}. \label{neq1}
\end{equation}

\noindent
When setting $p=0$ and $\nu=0$ in (\ref{neq1}) we get

\begin{equation}
  \frac{\partial {\vec U}}{\partial t}+ L\vec{U} = 
   {\vec f}. \label{neq2}
\end{equation}

\noindent
This equation describes Brownian motion, the seemingly random motion of a particle in a fluid caused by collisions with molecules in the fluid, Refs.~\onlinecite{coffey}-\onlinecite{zinn}. $L$ is the damping coefficient. The general solution of the equation of motion is

\begin{equation}
   \vec{U}(t) = \vec{U}(0)e^{-Lt} + \int_0^t dt' \vec{f}(t')e^{-(t-t')} .  \label{brownian}
\end{equation}

\noindent
We can add the pressure term to the equation (\ref{neq2}) 

\begin{equation}
  \frac{\partial {\vec U}}{\partial t}+ L\vec{U}- \frac{1}{\rho}\nabla p = 
   {\vec f}, \label{neq5}
\end{equation}

\noindent
and the solution is

\begin{equation}
   \vec{U}(t) = \vec{U}(0)e^{-Lt} + \int_0^t dt' [\vec{f}(t')e^{-(t-t')}  + \frac{1}{\rho}\nabla p].
\end{equation}
\\

\noindent
{\bf Brownian action} \\

Next, we look at the path-integral representation. 
The path-integral (\ref{acteff110000}) without the anomaly is

\begin{equation}
 S_{eff}[\nu=0] = \frac{1}{2} \int dtd^3 xd^3y \mu^i(t,{\vec x})\kappa({\vec x}-{\vec y})\mu_i(t,{\vec y}) -i\int
d^4x \mu^i(\partial_t u_i+Lu_i
-\partial_i \tilde{p}
) . \label{acteff110022}
\end{equation}

\noindent
We consider the determinant and use the result (\ref{eq:jac512})

\begin{eqnarray}
  \det | \partial_t+(e_j\partial_j)u+u(e_j\partial_j)-\nu\partial_j\partial_j | 
  &=& \frac{1}{2}L[u]-\nu F[u]
\label{eq:jac32121} \\ [1ex]
&=& {\rm constant} - \nu F[U] \\ [1ex]  
&=& 1 - \nu F[U]. 
\end{eqnarray}

\noindent
Here we have added a constant to the right-hand side of the determinant, which together with $L[u]/2$ forms the constant $1$. This addition applies only when $L[U]$ is constant and examining a change in $u$ in the determinant where

\begin{equation}
  \frac{\delta}{\delta u}  \det | \partial_t+(e_j\partial_j)u+u(e_j\partial_j)-\nu\partial_j\partial_j | 
  = -\nu\frac{\delta}{\delta u} F[u].
\end{equation}

\noindent
The action of the determinant is then

\begin{eqnarray}
   S_{D} &=&   - \vec{e} \int d^4x \ln (1-\nu F[u]) \\
   &\approx  & -\vec{e}
       \int d^4x  \Bigl ( (-\nu F[u])
       -\frac{1}{2}(-\nu F[u])^2 +... \Bigr ) 
   . \label{det20103}
\end{eqnarray}

\noindent
In (\ref{det20103}) we have used the expansion of the logarithmic function.  The path-integral can be written in the form

\begin{eqnarray}
 S_{eff} &=& \frac{1}{2} \int dtd^3 xd^3y \mu^i(t,{\vec x})\kappa({\vec x}-{\vec y})\mu_i(t,{\vec y}) -i\int
d^4x \mu^i(\partial_t u_i+L[u] u_i
-\partial_i \tilde{p} -\nu\partial_j\partial_j u_i
) \nonumber \\
&&-\vec{e} \int d^4x \ln (\frac{1}{2}L[u]-\nu F[u]) \\
&\approx & 
\frac{1}{2} \int dtd^3 xd^3y \mu^i(t,{\vec x})\kappa({\vec x}-{\vec y})\mu_i(t,{\vec y}) -i\int
d^4x \mu^i(\partial_t u_i+L[u] u_i
-\partial_i \tilde{p}) \nonumber \\
&& +\nu \int d^4x 
( i \mu^i \partial_j\partial_j u_i +  \vec{e}F[u]
) 
. \label{acteff1100}
\end{eqnarray}

\noindent
Here we have a fluid particle in random Brownian motion which is affected by the pump (first term) and the viscosity (last term).

\section{\label{sec5}Symmetry properties of $L$}

In this section, we investigate how the damping coefficient $L$ changes under the Navier-Stokes symmetries.
The stochastic Navier-Stokes equation (\ref{eq1}), $f=0$, is invariant with the following symmetries, Ref.~\onlinecite{cantwell}.\\

\begin{center}
\begin{tabular}{ l l  l  }
 a. &Space-translations & $\vec{U}(t,\vec{x}) =\vec{U}(t,\vec{x}+\vec{a}) \hspace*{0.5cm} \vec{a} \in  \mathbb{R}^3$ \\ [1ex]
 b. & Time-translations & $\vec{U}(t,\vec{x}) =\vec{U}(t+\tau,\vec{x}) \hspace*{0.5cm} \tau \in  \mathbb{R}$  \\  [1ex]
 c. & Space-reflections (parity) & $P\vec{U}(t,\vec{x}) =-\vec{U}(t,-\vec{x})$\\ [1ex]
 d. & Galilean transformations \hspace*{3cm}& $\vec{U}(t,\vec{x}) =\vec{U}(t,\vec{x}+\vec{a}t)-\vec{a} \hspace*{0.5cm} \vec{a} \in  \mathbb{R}^3$  \\ [1ex]
e. & Space-rotations & $\vec{U}(t,\vec{x}) = R^T\vec{U}(t,R\vec{x}) \hspace*{0.5cm} R \in  SO(3)$\\ [1ex]
f. & Scale invariance in space ($\nu=0$)& $ \vec{U}(t,\vec{x}) = \lambda^{h}\vec{U}(t,\lambda^{h}\vec{x}) \hspace*{0.5cm} \lambda \in  \mathbb{R}^+$ \\ [1ex]
g. & Scale invariance in time ($\nu=0$)& $ \vec{U}(t,\vec{x}) = \lambda^{-h}\vec{U}(\lambda^{h}t,\vec{x}) \hspace*{0.5cm} \lambda \in  \mathbb{R}^+$ \\ [1ex]
\end{tabular}
\end{center}

\noindent
Next, we consider how these symmetries affect $L$.\\

\noindent
{\bf a. Space-translations  } \\

The generator of the space translations is 

\begin{equation}
  \delta u_i = a_j\partial_j u_i,
\end{equation}

\noindent
where $a_j$ is constant.
The variation of the damping coefficient is then

\begin{equation}
  \delta L= e_j\partial_j\delta u_i e_i = e_k\partial_k(a_j\partial_j u_i)e_i = a_j\partial_j L =0,
\end{equation}

\noindent
and the $L$ is 

\begin{equation}
  \tilde{L}(t,\vec{x}) = L(t,\vec{x}+\vec{a}) = {\rm constant}. 
\end{equation}
\\

\noindent
{\bf b. Time-translations  } \\

In a same way, the generator of the time-translation is 

\begin{equation}
  \delta u_i = a \partial_t u_i,
\end{equation}

\noindent
and the $L$ is 

\begin{equation}
  \tilde{L}(t,\vec{x}) = L(t+a,\vec{x}) = {\rm constant}. 
\end{equation}
\\

\noindent
{\bf c. Parity  } \\

Parity cannot be represented by a sequence of infinitesimal transformations. It's easy to see though

\begin{equation}
  \tilde{L}(t,\vec{x}) = L(t,-\vec{x}) = {\rm constant}. 
\end{equation}
 \\

\noindent
{\bf d. Galilean symmetry  } \\

The infinitesimal change is

\begin{eqnarray}
  \delta u_i &=& a_jt\partial_j u_i-a_i, \\
  \delta \mu_i &=& a_jt\partial_j\mu_i, \\
   \delta \Psi_i &=& a_jt\partial_j\Psi_i,  \\
   \delta \bar{\Psi}_i &=& a_jt\partial_j\bar{\Psi}_i, 
\end{eqnarray}

\noindent
and the damping coefficient gives

\begin{equation}
   \delta L = e_j\partial_j \delta u_ie_i = at\partial_t L_ie_i= 0.
\end{equation}

\noindent
This corresponds to 

\begin{eqnarray}
   \tilde{L} &=& L(t,\vec{x}+\vec{a}t) = {\rm constant} ,\\
   \tilde{\vec{u}} &=& \vec{u}(t,\vec{x}+\vec{a}t)-\vec{a}. 
\end{eqnarray}

\noindent
{\bf e. Space-rotations  } \\

The field variations of the space-rotations are

\begin{eqnarray}
  \delta u_i &=& a(q_{kl}x_l\partial_k u_i-q_{ik}u_k),  \label{rot1}\\
  \delta \mu_i &=& a(q_{kl}x_l\partial_k \mu_i+q_{ki}\mu_k), \\
  \delta \Psi_i &=& a(q_{kl}x_l\partial_k \Psi_i-q_{ik}\Psi_k), \\
  \delta \bar{\Psi}_i &=& a(q_{kl}x_l\partial_k \bar{\Psi}_i+q_{ki}\bar{\Psi}_k), 
\end{eqnarray}

\noindent
where $q_{ij}$ is the rotation in $SO(3)$.
The change in $L$ is

\begin{equation}
  \delta L = 
    (R\cdot\nabla) 
    L = 0. \label{consu6} 
\end{equation}

\noindent
This means that the damping coefficient is invariant under the rotational symmetry, 

\begin{equation}
  \tilde{L}(t,\vec{x}) = L(t,R\vec{x}) = {\rm constant}.
\end{equation}

\noindent
{\bf f. Scale invariance in space ($\nu = 0$) } \\

The transformation rules are

\begin{eqnarray}
  \delta u_i &=& a (x_j \partial_j u_i- u_i),
\label{eq:deltau2} \\
  \delta \mu_i &=& a(x_j\partial_j \mu_i+2\mu_i),\label{eq:deltapsi2} \\
   \delta\tilde{p} &=& a(x_j\partial_j \tilde{p}-2\tilde{p}), \\
   \delta \Psi_i &=& a (x_j \partial_j \Psi_i+\frac{1}{2}\Psi_i), \\
   \delta \bar{\Psi}_i &=& a (x_j \partial_j \bar{\Psi}_i+\frac{1}{2}\bar{\Psi}_i).
\end{eqnarray}

\noindent
The change in $L$ is

\begin{equation}
  \tilde{L}(t,\vec{x}) = L(t,\lambda^h\vec{x}) = {\rm constant}.
\end{equation}

\noindent
{\bf g. Scale invariance in time  ($\nu = 0$)} \\

The transformation rules are

\begin{eqnarray}
\delta u_i &=& a(t\partial_t u_i + u_i),
\label{eq:deltau3} \\
  \delta \mu_i &=& a( t\partial_t \mu_i -\mu_i),\label{eq:deltapsi3} \\
   \delta \tilde{p} &=& a(t\partial_t \tilde{p} +2\tilde{p}), \\
  \delta \Psi_i &=& at\partial_t \Psi_i,  \\
   \delta \bar{\Psi}_i &=& at\partial_t \bar{\Psi}_i.
\end{eqnarray}

\noindent
The change in $L$ is

\begin{equation}
  \tilde{L}(t,\vec{x}) = \lambda^{-h}L(\lambda^h t,\vec{x}) = {\rm constant}.
\end{equation}

\noindent
We have shown that the damping coefficient obeys all the same symmetries that the stochastic Navier-Stokes equation has. In all transformations, it is constant, as required by the constraint. The symmetry transformations do not change the coefficient expect in the scale invariance in time. This is a feature with implications.\\

\noindent
{\bf Symmetries of the determinant actions} \\

In this section, we show that $L[u]$ must be constant due to the symmetries of the stochastic Navier-Stokes equation. First, suppose that $L[u]$ is not constant.

We start by applying symmetry transformations to the anomaly of the action, equation (\ref{det20100023}),

\begin{equation}
   S_{D1}(\nu = 0) =   - \vec{e} \int d^4x \bar{\Psi}\frac{1}{2}(e_j\partial_j )u(t,\vec{x})\Psi. \label{det201000234}
\end{equation}

\noindent
As can be verified, using the variations of the fields $u$, $\Psi$ and $\bar{\Psi}$, the action (\ref{det201000234}) is invariant under all local symmetry transformations a.-g.. In other words, for these local symmetries

\begin{equation}
   \delta S_{D1}(\nu = 0)_{\rm a.-g.}=  0,\label{det2010002341}
\end{equation}

\noindent
and the result is independent of whether $L[u]$ is constant. 
We then write (\ref{det201000234}) as

\begin{equation}
   S_{D1}(\nu = 0) 
   =  - \frac{1}{2}\vec{e} \int d^4x \bar{\Psi}L[u]\Psi. \label{det201000235}
\end{equation}

\noindent
The variation of (\ref{det201000235}) is 

\begin{eqnarray}
   \delta S_{D1}(\nu = 0) 
   &=&  - \frac{1}{2}\vec{e} \int d^4x (\delta\bar{\Psi}L[u]\Psi +
   \bar{\Psi}\delta L[u]\Psi +
   \bar{\Psi}L[u]\delta \Psi) \label{det201000236} \\
    &=& - \frac{1}{2}\vec{e} \int d^4x ( \bar{\Psi}\delta L[u]\Psi  + L[u](\delta\bar{\Psi}\Psi +
   \bar{\Psi}\delta \Psi)).
\end{eqnarray}

\noindent
We need and we have an additional condition.
Now looking at the eigenvalue equation (\ref{eigeneq}), we find that

\begin{equation}
   \Psi = \bar{\Psi},
\end{equation}

\noindent
and because the $\Psi$ and $\bar{\Psi}$ are anticommuting fields, the result is

\begin{equation}
   \delta S_{D1}(\nu = 0) 
    = - \frac{1}{2}\vec{e} \int d^4x  \bar{\Psi}\delta L[u]\Psi. \label{deltal}
\end{equation}

\noindent
Variation (\ref{deltal}) is zero only if $\delta L[u] = 0$ or $L[u]$ is constant.

We derive the same result in a different way. We have also another formula for the anomalous action, equation (\ref{det201000}),

\begin{equation}
   S_{D2}(\nu = 0) =    -\vec{e} \int d^4x \ln (\frac{1}{2}e_j\partial_j u(t,\vec{x})). \label{det2010001}
\end{equation}

\noindent
The Jacobi fields are missing from this action which enabled the vanishing action in field variations, (\ref{det201000234}) and (\ref{det2010002341}). However,
for consistency reasons, for the same local symmetries a.-g., the variation of the action (\ref{det2010001}) must also vanish under the same symmetry transformations, i.e.

\begin{equation}
   \delta S_{D2}(\nu = 0)_{\rm a.-g.} = 0.
\end{equation}

\noindent
Therefore, the constraint condition 

\begin{equation}
 L[u]
  = {\rm constant},
\label{eq:jac321255}
\end{equation}

\noindent
is the only valid choice of gauge that the stochastic Navier-Stokes turbulence equation can have when examining the symmetries of this equation. Otherwise, the symmetry properties of the anomalous actions (\ref{det201000234}) and (\ref{det2010001}) are different.

In summary, we have shown in two different ways that $L[u]$ must be constant.

\section{\label{sec55}Vortices and monopoles}

A Vortex is a stable time-independent solution to a set of classical field equations that has finite energy in two spatial dimensions; it is a two-dimensional soliton. In three spatial dimensions, a vortex becomes a string, a classical solution with finite energy per unit length, Ref.~\onlinecite{preskill}.

In this section, we examine $L$ for a Brownian fluid particle solution, (\ref{brownian}),

\begin{equation}
   \vec{U}(t,\vec{x}) = \vec{U}(0,\vec{x})e^{-Lt } + \int_0^t dt' \vec{f}(t',\vec{x})e^{-(t-t')}. \label{brownian1}
\end{equation}

\noindent
The constraint $L$ becomes

\begin{equation}
    L \vec{e}= e_j \partial_j u \vec{e} = \int_{0}^{t} dt' e_j \partial_j \vec{f}(t',\vec{x})e^{-(t-t')} = {\rm constant}\cdot\vec{e}.\label{condis0}
\end{equation}

\noindent
We have assumed that the first term of (\ref{brownian1}) can be neglected in large values of time. Taking a derivative with respect to time we have

\begin{equation}
     e_j \partial_j f_i(t,\vec{x}) = 0.\label{condis1}
\end{equation}

\noindent
We assume a pulse which has a Gaussian shape

\begin{equation}
  f_i \sim \exp(-a_jx_j^2),
\end{equation}

\noindent
where $a_i$ is constant. Equation (\ref{condis1}) gives

\begin{equation}
  \sum_{j=1}^{3}a_je_jx_j = 0.  \label{condis2}
\end{equation}

\noindent
The condition (\ref{condis2}) create a constraint on the shape of the soliton. In two dimensions, the vortex has a Gaussian shape and it is rotationally symmetric in $u$. In three spatial dimensions, a vortex has a shape

\begin{equation}
  f_i \sim \exp(-a_1x_1^2-a_2x_2^2-[(-a_1e_1x_1-a_2e_2x_2)/(a_3e_3)]^2),
\end{equation}

\noindent
and this is a string.

There are two things to note. First, the stochastic force $f$ is the source of the anomaly and the vortex. Without a force there are no vortices. And conversely, when a stochastic force is present, there is an anomaly or vortex in the theory. Second, as considered earlier in this article,

\begin{equation}
  \vec{\mu}\cdot(\vec{U}\cdot\nabla)\vec{U}
\end{equation}

\noindent
indeed describes the interaction term between the field and the vortex.

\section{\label{sec6}Brownian particle in scaling symmetry}

As a more concrete example, we consider the following local time reparameterization, $\beta = \beta(t)$, $a$ and $b$ are constants,

\begin{eqnarray}
  \tilde{t} &=& \beta(t)^a, \label{local1}\\
  \tilde{\vec{x}} &=& \vec{x}\beta'(t)^b ,\\
  \tilde{u}_i &=& u_i\beta'(t)^{a-b} ,\\
  \tilde{\mu}_i &=& \mu_i\beta'(t)^{2b-a}, \label{local2}\\
  \tilde{\tilde{p}} &=& \tilde{p}\beta'(t)^{2a-2b}.
\end{eqnarray} 

\noindent
This combines the scaling of time and space symmetries.
The transformation corresponds to the following field variations

\begin{eqnarray}
  \delta u_i &=& a\beta \partial_t u_i +b\beta' x_j \partial_j u_i+(a-b)\beta' u_i,
\label{eq:deltau} \\
  \delta \mu_i &=& a\beta \partial_t \mu_i +b\beta' x_j\partial_j \mu_i+(2b-a)\beta'\mu_i,\label{eq:deltapsi} \\
  \delta \tilde{p} &=& a\beta \partial_t \tilde{p} +b\beta' x_j\partial_j \tilde{p}+(2a-2b)\beta'\tilde{p}.
\end{eqnarray}

\noindent
The change in the action is

\begin{eqnarray}
  \frac{\delta S_{eff}}{b} &=& \frac{3h-1}{2}\int dtd^3xd^3y \beta' \mu^i(t,\vec{x})\kappa(\vec{x}-\vec{y})\mu_i(t,\vec{y}) 
- i  \int dx^4 \beta'' \mu^i (x_j \partial_j u_i-hu_i) \nonumber \\
&&
+ i (h+1)\nu \int dx^4 \beta'\mu^i \partial_j\partial_j u_i  -
\vec{e}\frac{1}{b}\delta \int dx^4 \Bigl (\frac{1}{2}L[u] - \nu  F[u]  \Bigr ) ,
\label{eq:final0000}
\end{eqnarray}

\noindent
where $h=(b-a)/b$
The second term vanishes either if

\begin{eqnarray}
 &{\rm a)}&\beta'' = 0, \hspace*{2cm}{\rm or}  \label{condi1} \\
 &{\rm b)}&x_j\partial_ju_i - hu_i = C/\beta'', \label{condi2}
\end{eqnarray}

\noindent
where $C$ is constant. In case b) the integral becomes zero based on conservation of motion of the center of mass. \\

\noindent
{\bf Brownian particle} \\

We set $\beta = t$. Then the change is

\begin{equation}
  \tilde{u}_i = \lambda^{a-b}u_i (\lambda^at,\lambda^b\vec{x}),\label{local12}
\end{equation} 

\noindent
and

\begin{eqnarray}
  \delta u_i &=& at \partial_t u_i +b x_j \partial_j u_i+(a-b)u_i,
\label{eq:deltauu} \\
  \delta \mu_i &=& at \partial_t \mu_i +b x_j\partial_j \mu_i+(2b-a)\mu_i,\label{eq:deltapsiu} \\
  \delta \tilde{p} &=& at \partial_t \tilde{p} +bx_j\partial_j \tilde{p}+(2a-2b)\tilde{p}, \\
  \delta \tilde{L} &=& at \partial_t L +bx_j\partial_j L+aL=0.
\end{eqnarray}

\noindent
This is a Brownian fluid particle in the presence of the stochastic pump and the viscosity

\begin{equation}
   \vec{U}(t) = \vec{U}(0)e^{-L[u]t} + \int_0^t dt' [ \vec{g}(t')e^{-(t-t')}  +\frac{1}{\rho}\nabla p], \label{brow1}
\end{equation}

\noindent
and the change of the damping coefficient is

\begin{equation}
  \tilde{L} = \lambda^aL(\lambda^at,\lambda^b\vec{x}). \label{bL}
\end{equation}

\noindent
The force $\vec{g}$ is obtained as an equation of motion from the equation (\ref{eq:S}), $\nu=0$,

\begin{equation}
  \frac{\delta S}{\delta \mu} = \vec{g}(t) - (\partial_t \vec{U}+(\vec{U}\cdot\nabla)\vec{U}-\nabla \tilde{p})=0,
\end{equation}

\noindent
and

\begin{equation}
  \vec{g}(t) = \frac{1}{2} \int d^3xd^3y \Bigl ( \vec{\mu}(t,\vec{x})\kappa(\vec{x}-\vec{y}) 
  + \kappa(\vec{x}-\vec{y})\vec{\mu}(t,\vec{y}) \Bigr ) .
\end{equation}

\noindent
We have also used the Wick rotation, $\vec{\mu} \to i\vec{\mu}$.
Based on (\ref{bL}) the solution (\ref{brow1}) scales in time but not in space.
The symmetry is visible when the pump and the viscosity are in balance

\begin{equation}
  \frac{\delta S_{eff}}{b} \approx \frac{3h-1}{2}\int dtd^3xd^3y  \mu^i(t,\vec{x})\kappa(\vec{x}-\vec{y})\mu_i(t,\vec{y}) 
+ \nu \int d^4x 
( i (h+1)\mu^i \partial_j\partial_j u_i +  \vec{e}\frac{1}{b}\delta F[u] 
)  = 0.
\label{eq:final000000}
\end{equation}

\noindent
We have used the representation of the Brownian random motion (\ref{acteff1100}). The pump term injects the energy and the viscosity dissipates it.\\

\section{\label{sec7}Conclusions}

Navier-Stokes turbulence is a chaotic and turbulent flow of a fluid characterized by rapid and irregular movement of fluid particles. On the other hand, Brownian motion is the random movement of particles suspended in a liquid, which is caused by their collision with fast-moving atoms or molecules in the surrounding medium. Random Brownian motion in Navier-Stokes turbulence represents a complex interaction between two separate phenomena in fluid dynamics and statistical physics. The connection between Brownian motion and Navier-Stokes turbulence arises when looking at small particles or tracer particles suspended in a turbulent fluid.

In this article, we have shown that there is an anomaly in the Navier-Stokes turbulence driven by the Gaussian random force. The anomaly must somehow be handled in stochastic turbulence calculations. If no choice is made, an inherent feature of the anomaly is lost. Therefore, our result suggests a third law for the original Navier-Stokes equations (\ref{eq1})-(\ref{eq1b})

\begin{eqnarray}
  {\rm A)} && \frac{\partial {\vec U}}{\partial t}+ ({\vec U} \cdot \nabla) {\vec U}  = 
   \frac{1}{\rho}\nabla p
  +\nu \nabla^2 {\vec U} +{\vec f}, \label{eq1a}\\ [1ex]
  {\rm B)} && \nabla\cdot \vec{U} = 0, \label{eq1bb} \\ [1ex]
  {\rm C)} && {\rm Rule\hspace*{1mm}how \hspace*{1mm}to \hspace*{1mm}treat\hspace*{1mm} the\hspace*{1mm} anomaly \hspace*{1mm}(=fixing \hspace*{1mm}the \hspace*{1mm}gauge).}\label{eq1bc}
\end{eqnarray}

\noindent
We need a rule, Eq.~(\ref{eq1bc}), that defines how to fix the "gauge" for the velocity field. 
In this article we selected the gauge that adds a constant constraint  

\begin{eqnarray}
  {\rm A)} && \frac{\partial {\vec U}}{\partial t}+ ({\vec U} \cdot \nabla) {\vec U}  = 
   \frac{1}{\rho}\nabla p
  +\nu \nabla^2 {\vec U} +{\vec f}, \label{eq2a}\\ [1ex]
  {\rm B)} && \nabla\cdot \vec{U} = 0, \label{eq2bb} \\ [1ex]
  {\rm C)} && (\vec{e}\cdot\nabla)\vec{U} = {\rm constant}.\label{eq2bc}
\end{eqnarray}

\noindent
This choice of gauge, Eq.~(\ref{eq2bc}), forces the anomaly to be a number and removes the anomalous term from the theory. Instead, a monopoly-type object is formed, whose "charge" is "constant". 
If we want to examine the local symmetries of the stochastic Navier-Stokes equation, this gauge is the only correct choice.
The gauge (\ref{eq2bc}) is consistent with the Brownian damping coefficient of a random fluid particle. The difference to the traditional Brownian factor is that in this gauge selection the coefficient depends on the velocity field and it also follows all the symmetries of the Navier-Stokes equation. Another selection of the gauge is also possible, and the choice may depend on the physical problem under investigation.

\end{document}